\documentclass[a4paper,keywords,synopsis]{iucr}
\usepackage[latin9]{inputenc}
\usepackage{textcomp}
\usepackage{amsmath}
\usepackage{amsbsy}
\usepackage{graphicx}
\usepackage{esint}
\makeatletter

\usepackage{upgreek}

     \paperprodcode{a000000}      
     \paperref{xx9999}            
     \papertype{FA}               

     \paperlang{english}          
     \journalcode{J}             
     \journalyr{2019}
     \journalreceived{\relax}
     \journalaccepted{\relax}
     \journalonline{\relax}

\makeatother

\begin{document}
\title{X-ray diffraction from strongly bent crystals and spectroscopy of
XFEL pulses}
\shorttitle{Diffraction from strongly bent crystals}
\author[a]{Vladimir M.}{Kaganer}
\author[b]{Ilia}{Petrov}
\author[b]{Liubov}{Samoylova}
\aff[a]{Paul-Drude-Institut für Festkörperelektronik, Leibniz-Institut im
Forschungsverbund Berlin e.\,V., Hausvogteiplatz 5--7, 10117 Berlin,
Germany}
\aff[b]{European XFEL GmbH, Holzkoppel 4, 22869 Schenefeld, Germany}
\shortauthor{Kaganer, Petrov, and Samoylova}
\keyword{x-ray free-electron lasers}
\keyword{x-ray spectroscopy}
\keyword{bent crystals}
\keyword{diamond crystal optics}
\keyword{femtosecond x-ray diffraction}
\keyword{dynamical diffraction}
\maketitle
\begin{synopsis}
Strongly bent crystal diffracts kinematically when the bending radius
is small compared to the critical radius given by the ratio of the
extinction length to the Darwin width of the reflection. Under these
conditions, the spectral resolution of the XFEL pulse is limited by
the crystal thickness and can be better than under dynamical diffraction
conditions. 
\end{synopsis}
\begin{abstract}
The use of strongly bent crystals in spectrometers for pulses of a
hard x-ray free-electron laser is explored theoretically. Diffraction
is calculated in both dynamical and kinematical theories. It is shown
that diffraction can be treated kinematically when the bending radius
is small compared to the critical radius given by the ratio of the
Bragg-case extinction length for the actual reflection to the Darwin
width of this reflection. As a result, the spectral resolution is
limited by the crystal thickness, rather than the extinction length,
and can become better than the resolution of a planar dynamically
diffracting crystal. As an example, we demonstrate that spectra of
the 12~keV pulses can be resolved in 440 reflection from a 20~\textmu m
thick diamond crystal bent to a radius of 10~cm.
\end{abstract}

\section{Introduction}

Bent single crystals are commonly used as the x-ray optic elements
for beam conditioning as well as the analyzers for x-ray spectroscopy.
The dynamical diffraction from bent crystals has been a topic of numerous
studies over decades \cite{penning61,kato64,bonse64,chukhovskii77,chukhovskii78,kalman83,gronkowski84,chukhovskii89,gronkowski91,honkanen18}.

Recently, hard x-ray free-electron lasers (XFELs) went into operation
around the world \cite{LCLS2010,SACLA2012,SwissFEL2017,PAL-XFEL,EuXFEL2017}.
At all of these sources, XFEL pulses originate from random current
fluctuations in the electron bunch \cite{saldin:book}, which gives
rise to an individual time structure and energy of each pulse. The
energy spectra of single pulses need to be characterized in a noninvasive
way, allowing further use of the same pulses in the experiments.

Two basic requirements for the spectrometers---the acceptance range
of photon energy and the energy resolution---follow from the duration
of the pulse and the duration of the spikes in it \cite{saldin:book}.
A spike duration of $\tau_{\mathrm{s}}=0.1$~fs gives rise to an
energy range that needs to be covered by the spectrometer $\Delta E=h/\tau_{\mathrm{s}}=40$~eV,
where $h=4.13$~eV$\cdot$fs is the Planck constant. When an x-ray
beam of a width $w$ is incident on a crystal bent to a radius $R$,
the range of available Bragg angles $w/R$ has to exceed the required
angular range $\Delta\theta=(\Delta E/E)\tan\theta_{B}$, where $\theta_{B}$
is the Bragg angle. Taking $\tan\theta_{B}=1$ for simplicity and
$E=12$~keV as a reference energy, we find that, for a beam of a
width $w=500$~\textmu m, the curvature radius should be less than
$R=15$~cm to cover whole spectrum. The bending radii of 5~cm for
a 10~\textmu m thick silicon crystal \cite{zhu12} and 6~cm for
a 20~\textmu m thick diamond \cite{boesenberg17} are reached. The
resolution requirement for a spectrometer follows from the total duration
of a pulse up to $\tau_{\mathrm{p}}=50$~fs, which gives $\delta E=h/\tau_{\mathrm{p}}=0.08$~eV.

Different types of spectrometers based on silicon crystals have been
proposed, built, and tested for this purpose. They employ a focusing
mirror with a flat diffracting crystal \cite{yabashi06,inubushi12},
a focusing grating \cite{karvinen12}, a bent diffracting crystal
\cite{zhu12}, and a flat grating with a bent diffracting crystal
\cite{makita15}. Recently, a spectrometer based on a bent thin diamond
crystal has been designed and tested \cite{boesenberg17,samoylova19}
for high repetition rate XFEL sources, such as the European XFEL and
LCLS II. The diamond is the material of choice for high repetition
rate XFELs because only diamond can sustain the enormous peak heat
load and prevent severe vibrations when thermal stress wave is excited
under repeated heat load in the megahertz range at a resonant frequency
of the thin crystal plate.

The studies of the XFEL pulses using diffraction on bent crystals
\cite{zhu12,makita15,boesenberg17,rehanek17} treated diffraction
purely geometrically, as a mirror reflection of a geometric ray at
a point where it meets the crystal surface. The process of diffraction
in the crystal has not been taken into account, despite the crystal
thicknesses of 10 to 20~\textmu m, which exceed the extinction lengths
of dynamical diffraction for respective reflections (see estimates
in the next section).

The studies of dynamical diffraction on bent crystals cited above
considered the bending of thick crystals to radii varying from hundreds
of meters to single meters. The curvature radius of some hundreds
of meters already provides detectable broadening of the Darwin rocking
curve, while the bending to a radius of one meter strongly modifies
it. The results of these studies are not applicable to the case under
consideration, where the crystal is thin and the bending radius is
much smaller.

In the present paper, we consider x-ray diffraction on crystals bent
to a radius of 10~cm or less. In case of such strong bending, the
incident x-ray wave remains at diffraction conditions (i.e., within
the Darwin width of the actual reflection) only when propagating through
distances small compared to the extinction length. As a result, a
back scattering of the diffracted wave to the transmitted one is minor
and diffraction is kinematical. We calculate diffraction from a bent
crystal in both dynamical and kinematical theories and establish the
applicability criterion for the approximation of kinematical diffraction.

We obtain a displacement field in the bent crystal by considering
cylindrical bending of an elastically anisotropic rectangular thin
plate by two momenta applied to its orthogonal edges. We show that,
for a 110 oriented diamond plate, the elastic constants of diamond
give rise to a very small strain variation along plate normal because
the Poisson effect on bending is almost completely compensated by
the effect of anisotropy. As a result, the resolution of a bent-crystal
spectrometer is limited by the crystal thickness and can be better
than the resolution of a non-bent crystal, limited by the extinction
length.

We simulate XFEL spectra after diffraction on a bent crystal and show
that an energy resolution of $3\times10^{-6}$, or 0.04~eV for the
x-ray energy of 12~keV, can be reached on diffraction on a 20~\textmu m
thick diamond crystal bent to a radius of 10~cm. We also take into
account the free-space propagation of the waves diffracted by the
bent crystal to the detector (Fresnel diffraction) and describe modifications
of the spectra due to a finite distance to the detector.

\section{Dynamical vs. kinematical diffracted intensities}

For numerical estimates in this section, we consider, as a reference
example, symmetric Bragg reflection 440 of the x-rays with the energy
$E=12$~keV (wavelength $\lambda=1.03$~Å) from a $D=20$~\textmu m
thick diamond crystal bent to a radius $R=10$~cm.

When the crystal is not bent and oriented to satisfy the exact Bragg
condition in symmetric reflection geometry, penetration of an x-ray
wave in it is governed by the extinction length $\Lambda$, defined
as a depth at which the amplitude of the wave decreases by a factor
of $e$ (correspondingly, intensity decreases $e^{2}$ times). The
extinction length is equal to $\Lambda=\lambda\sin\theta_{B}/\pi\sqrt{\left|\chi_{h}\chi_{\bar{h}}\right|}$,
where $\chi_{h}$ and $\chi_{\bar{h}}$ are the Fourier components
of crystal susceptibility. For our example, the extinction length
amounts to \cite{stepanov:www} $\Lambda=13.6$~\textmu m. The crystal
thickness in our reference example is larger than the extinction length,
and hence diffraction in a non-bent crystal should be calculated in
the framework of dynamical diffraction theory.

Dynamical diffraction (strong coupling between the transmitted and
the diffracted waves) takes place as long as the lattice distortions
(the lattice spacing and the orientation of lattice planes) do not
change on the distance $\Lambda$, or the change is much less than
the width of the Darwin curve $\Delta\theta_{B}=2\sqrt{\left|\chi_{h}\chi_{\bar{h}}\right|}/\sin2\theta_{B}$,
which in our case is $\Delta\theta_{B}=4.2$~\textmu rad \cite{stepanov:www}.
For a bent crystal of radius $R$, the gradient of distortions is
$1/R$ and its change on the distance of the extinction length is
$\Lambda/R$. If the crystal is so strongly bent that this change
is much larger than $\Delta\theta_{B}$, dynamical diffraction effects
become negligible, since the path of the transmitted wave under diffraction
conditions occurs much smaller than the extinction length. Such an
estimate is similar to the treatment of the interbranch scattering
in the vicinity of crystal lattice defects by \citeasnoun{authier70aca}
and \citeasnoun{authier70pss} and predicts that the dynamical diffraction
effects can be neglected for bending radii $R\ll R_{c}$, where 
\begin{equation}
R_{c}=\Lambda/\Delta\theta_{B}=\left(\Lambda^{2}Q/4\right)\cot\theta_{B}.\label{eq:D1}
\end{equation}
Here $Q=(4\pi/\lambda)\sin\theta_{B}$ is the diffraction vector.
For our example, $R_{c}=3.2$~m.

To verify the applicability of the approximation of kinematical diffraction,
we perform calculations of Bragg diffraction from a bent crystal plate
in both dynamical and kinematical diffraction theories. In the calculations,
the Fourier component of susceptibility $\chi_{h}$ can be varied
arbitrarily. The kinematical scattering amplitude is proportional
to $\chi_{h}$ (and hence intensity is proportional to $\left|\chi_{h}\right|^{2}$)
for any fixed bending radius, while the dynamical scattering amplitude
depends on both $\chi_{h}$ and $R$ in a complicated way. Hence,
the applicability of the kinematical theory can be established in
the framework of dynamical diffraction, by studying the dependence
of the diffracted intensity on $\chi_{h}$. This is done in the present
section. In the next section, we directly compare the kinematical
and the dynamical scattering intensities.

Dynamical diffraction is calculated by numerical solution of the Takagi-Taupin
equations \cite{takagi62,takagi69,taupin64} 
\begin{equation}
\frac{\partial\mathcal{E}_{0}}{\partial s_{0}}=\frac{i\pi\chi_{\bar{h}}}{\lambda}e^{i\mathbf{Q}\cdot\mathbf{u}}\mathcal{E}_{h},\qquad\frac{\partial\mathcal{E}_{h}}{\partial s_{h}}=\frac{i\pi\chi_{h}}{\lambda}e^{-i\mathbf{Q}\cdot\mathbf{u}}\mathcal{E}_{0}.\label{eq:D1a}
\end{equation}
Here $\mathcal{E}_{0}$ and $\mathcal{E}_{h}$ are the amplitudes
of the transmitted and the diffracted waves, $s_{0}$ and $s_{h}$
are the coordinates in the propagation directions of these waves,
$\mathbf{Q}$ is the scattering vector, and $\mathbf{u}(\mathbf{r})$
is the displacement vector. It describes displacement of atoms from
their positions in a reference non-bent crystal. The displacement
$\mathbf{u}(\mathbf{r})$ changes the susceptibility $\chi(\mathbf{r})$
of the reference crystal to $\chi(\mathbf{r}-\mathbf{u}(\mathbf{r}))$,
and Fourier expansion of the susceptibility over reciprocal lattice
vectors $\mathbf{Q}$ gives rise to the terms $\exp(\pm i\mathbf{Q}\cdot\mathbf{u}(\mathbf{r}))$
in equations (\ref{eq:D1a}). The algorithm of numerical solution
of Eqs.~(\ref{eq:D1a}) was proposed by \citeasnoun{authier68}
and revisited later by \citeasnoun{gronkowski91} and \citeasnoun{shabalin17}.
To proceed to numerical solution of the Takagi-Taupin equations, we
specify first the diffraction geometry and the displacement field
$\mathbf{u}(\mathbf{r})$ entering these equations.

\begin{figure}
\includegraphics[width=7cm]{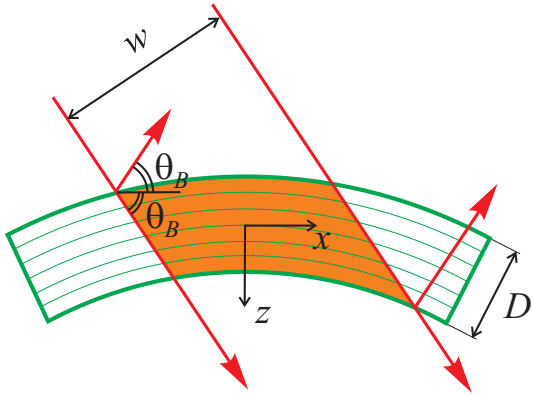}
\caption{Geometry of symmetric Bragg diffraction from a bent crystal.}
\label{fig:geometry} 
\end{figure}

Figure \ref{fig:geometry} sketches symmetric Bragg diffraction from
a bent crystal plate. The scattering plane is the $xz$ plane, and
the crystal is bent about $y$ axis. An ultrashort XFEL pulse, represented
by its energy spectrum, is a coherent superposition of the waves with
the same propagation direction and different wavelengths. We take
a reference wavelength in the middle of the pulse spectrum and choose
the origin $(x=0,z=0)$ at a point in the middle plane of the crystal
plate where the incident and the diffracted waves of the reference
wavelength make the same angle $\theta_{B}$ with the lattice planes.

The incident beam is restricted by a width $w$. The width of the
wavefront of an XFEL pulse at the experiment is about 1~mm, much
larger than the crystal thickness, but it can be focused to tens of
microns, comparable with the crystal thickness. The estimate below
shows that, if the beam is not focused, its width is much larger than
the width of the diffracting region of the strongly bent crystal.
The outer parts of the beam occur out of Bragg diffraction, and hence
the beam width does not restrict diffraction.

Besides a focused incident beam, the width of the incident beam becomes
essential when the bent crystal is rotated to measure its rocking
curve \cite{samoylova19}. The diffracted intensity decreases when the crystal is rotated
such that the region of the crystal oriented at the Bragg angle to
the incident beam goes out of the illuminated region of the crystal.
This is reached for the angular deviations from the Bragg angle $\delta\theta\sim w/R$.
Hence, the width of the rocking curve of a bent crystal is given by
the width of the incident beam. In all other situations, i.e., if the
incident beam is not focused to a few tens of microns at the crystal
and the angular deviation of the crystal is small compared with its
rocking curve width, the width of the incident beam is irrelevant.
In the practical case, we take $w=500$~\textmu m in the calculations
below and ensure that the diffracted intensity does not change with
a further increase of the beam width.

In symmetric Bragg case diffraction considered here, the diffraction
vector $\mathbf{Q}$ is in the negative direction of $z$ axis and
$\mathbf{Q}\cdot\mathbf{u}=-Qu_{z}$, so that only $z$-component
of the displacement vector in the bent crystal is of interest. It
is calculated in Appendix \ref{sec:AppendixA} taking into account
the elastic anisotropy of a crystal with cubic symmetry. The displacement
field in a crystal cylindrically bent to a radius $R$ can be written
as {[}cf.~Eq.~(\ref{eq:A8}){]} 
\begin{equation}
u_{z}=(x^{2}+\alpha z^{2})/2R,\label{eq:D2}
\end{equation}
where the constant $\alpha$ depends on the elastic moduli and the
crystal orientation {[}see Eq.~(\ref{eq:A9}){]}. The elastic moduli
of diamond give rise to exceptionally small values of $\alpha$: we
find $\alpha=0.02$ for an $110$ oriented plate bent about $001$
axis and $\alpha=0.047$ for an $111$ oriented plate bent about $11\bar{2}$
axis. For comparison, the elastic moduli of silicon result in $\alpha=0.18$
and 0.22 for these two orientations.

\begin{figure}
\includegraphics[width=\columnwidth]{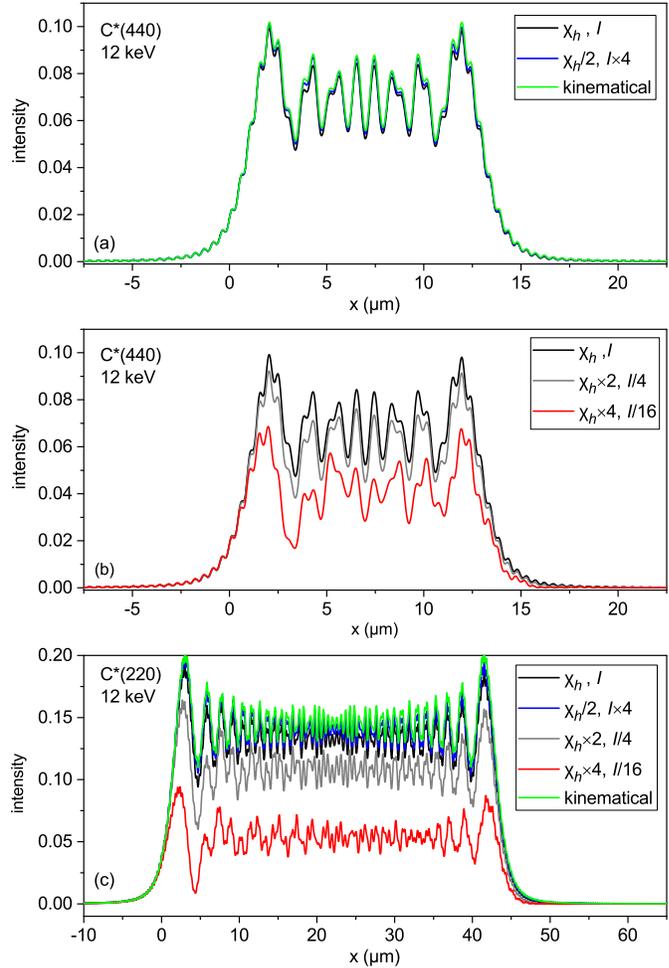}
\caption{Dynamical and kinematical intensities of diffracted wave at the crystal
surface in symmetric Bragg reflections (a,b) 440 and (c) 220 from
20 \textmu m thick diamond crystal bent to a radius of 10~cm. The
x-ray energy is 12 keV. Dynamical diffraction calculations
for the x-ray susceptibilities $\chi_{h}$ of the respective reflections
(black lines) are repeated taking susceptibility smaller by a factor
of 2, with the intensity multiplied by a factor of 4 (blue lines).
Dynamical diffraction calculations are also performed with the susceptibilities
$\chi_{h}$ multiplied by factors 2 and 4, and with the respective
intensities divided by factors 4 and 16 (gray and red lines). The
kinematical intensities calculated by Eq.~(\ref{eq:D4}) are shown
by green lines.}
\label{fig:dynkin} 
\end{figure}

Figure \ref{fig:dynkin}(a) shows by black line the intensity distribution
of the dynamically diffracted wave at the crystal surface for our
example case. The spatial width of the diffracted wave is much smaller
than the width of the incident wave and is determined by the crystal
thickness projected to the surface at the Bragg angle. The amplitude
of the incident wave is taken equal to 1. The amplitude of the diffracted
wave is small compared to it, which points out to the kinematical
diffraction.

To verify the kinematical nature of diffraction further, we perform
the same calculation but, instead of the susceptibility $\chi_{h}$,
use the value $\chi_{h}/2$ without changing any other parameter.
When the approximation of kinematical diffraction is applicable, the
diffracted amplitude is expected to be proportional to $\chi_{h}$,
so that the intensity is proportional to $|\chi_{h}|^{2}$. Hence,
we multiply the calculated intensity by a factor of 4 (blue line)
and compare with the former calculation with the initial value $\chi_{h}$
(black line). The curves practically coincide, which further evidences
the kinematical nature of diffraction. Thus, Fig.~\ref{fig:dynkin}(a)
approves, by means of the calculations made in the framework of dynamical
theory, the applicability of the approximation of kinematical diffraction
for curvature radii small compared with the critical radius (\ref{eq:D1}).

In Fig.~\ref{fig:dynkin}(b), we calculate dynamical diffraction
intensity in the same reflection but with the susceptibility $\chi_{h}$
increased by factors 2 and 4, with the aim to establish the applicability
limits of the approximation of kinematical diffraction. Since the
critical radius $R_{c}$ in Eq.~(\ref{eq:D1}) is proportional to
$|\chi_{h}|^{2}$, the increase of $\chi_{h}$ by a factor of 2 reduces
the critical radius from 3.2~m to 80~cm, still large compared with
the bending radius of 10~cm. The calculated curve {[}gray line in
Fig.~\ref{fig:dynkin}(b){]} deviates from the reference curve (black
line) mostly by a scale factor. When the susceptibility $\chi_{h}$
is increased by a factor of 4, and hence the critical radius reduced
to 20~cm, the calculated diffraction intensity (red curve) notably
differs from the reference black curve not only in scale but also
in the shape of fringes. Thus, approaching the critical radius (\ref{eq:D1})
results in a strong modification of the diffracted intensity.

Figure \ref{fig:dynkin}(c) collects similar calculations for C{*}(220)
reflection under the same conditions. For this reflection of 12~keV
x-rays, the Bragg case extinction length and the Darwin width are
\cite{stepanov:www} $\Lambda=4.17$~\textmu m and $\Delta\theta_{B}=8.63$~\textmu rad,
so that the critical radius $R_{c}=\Lambda/\Delta\theta_{B}=48$~cm,
and the bending radius of 10~cm occurs closer to the critical radius.
Calculation with the susceptibility $\chi_{h}$ for this reflection
(black line) and for 2 times smaller susceptibility (blue line) slightly
differ by a scale factor, so that the approximation of kinematical
diffraction is applicable but close to its applicability border. When
the susceptibility is increased by a factor of 2 (gray line), the
critical radius becomes 12~cm, close to the bending radius. The calculated
diffraction intensity notably differs from the reference black curve.
When the susceptibility is increased by a factor of 4 and the critical
radius becomes as small as 3~cm, the fringes of the calculated intensity
(red curve) do not follow the reference curve, again confirming that,
for the radii smaller than the critical radius (\ref{eq:D1}), the
use of dynamical theory is necessary.

\begin{figure}
\includegraphics[width=\columnwidth]{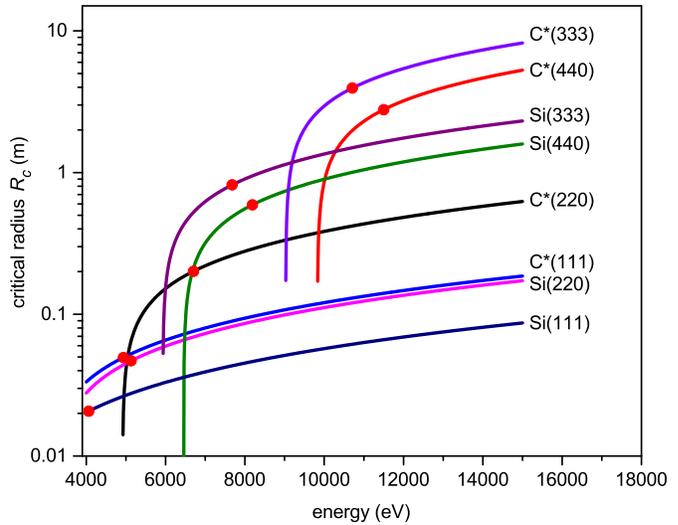}
\caption{The x-ray energy dependence of the critical radii given by Eq.~(\ref{eq:D1})
for several reflections of diamond and silicon. The point at each
curve marks an energy such that, for a crystal thickness 20~\textmu m
and distance to detector 1~m, the width of the beam diffracted from
bent crystal is equal to the width of the first Fresnel zone. The
Fraunhofer approximation is applicable, under these conditions, for
energies smaller than the marked energy (see Sec.~\ref{sec:FresnelDiffraction}
for details).}
\label{fig:Rc} 
\end{figure}

The analysis in the next sections shows that the applicability of
the approximation of kinematical diffraction not only simplifies calculation
of the intensity diffracted by the bent crystal but leads to a resolution
better than given by the Darwin width of dynamical diffraction. Therefore,
the critical radii for different reflections are of interest. Figure
\ref{fig:Rc} presents critical radii for symmetric Bragg reflections
from diamond and silicon crystals as a function of the x-ray energy.
Since the energy range presented in Fig\@.~\ref{fig:Rc} is far
from the absorption edges of carbon or silicon, the susceptibilities
$\chi_{h}$ are proportional to $\lambda^{2}$. Then, the extinction
length does not depend on $\lambda$ and, as it follows from the second
equality in Eq.~(\ref{eq:D1}), the energy dependence of the critical
radius is simply given by the factor $\cot\theta_{B}$.

Thus, in this section, we have verified, entirely by means of calculations
performed in the framework of dynamical theory, the criterion (\ref{eq:D1})
for applicability of the approximation of kinematical diffraction.
In the next section, we calculate the kinematical amplitude and compare
it with the calculations of dynamical diffraction.

\section{Kinematical diffraction amplitude at crystal surface and in far field}

\subsection{Amplitude at crystal surface}

The kinematical diffraction amplitude at the crystal surface $\mathcal{E}_{h}^{\mathrm{kin}}(x)$
can be obtained by neglecting the influence of the diffracted wave
$\mathcal{E}_{h}(\mathbf{r})$ on the transmitted wave $\mathcal{E}_{0}(\mathbf{r})$
in the first Takagi-Taupin equation (\ref{eq:D1a}). Then the amplitude
of the transmitted wave in the crystal is given by the first equation
shortened to $\partial\mathcal{E}_{0}/\partial s_{0}=0$, which gives
$\mathcal{E}_{0}(\mathbf{r})=1$. The diffracted wave is determined
by the solution of the second equation, which becomes now 
\begin{equation}
\frac{\partial\mathcal{E}_{h}}{\partial s_{h}}=\frac{i\pi\chi_{h}}{\lambda}e^{-i\mathbf{Q}\cdot\mathbf{u}},\label{eq:D2a}
\end{equation}
with the boundary condition $\mathcal{E}_{h}=0$ at the bottom surface
of the crystal $z=D/2$. To simplify calculations, we restrict ourselves
in this section to the case of 110 oriented diamond crystal with its
very small value of $\alpha$, and take $\alpha=0$ in Eq.~(\ref{eq:D2}).
The general form of the kinematical integral is introduced and studied
in Sec.~\ref{sec:SpectralResolution}. The amplitude of the diffracted
wave at the top surface $z=-D/2$ is 
\begin{equation}
\mathcal{E}_{h}^{\mathrm{kin}}(x)=i\frac{\pi\chi_{h}}{\lambda\cos\theta_{B}}\intop_{x-D\cot\theta_{B}}^{x}\exp\left(i\frac{Qx'^{2}}{2R}\right)\,dx'.\label{eq:D3}
\end{equation}
The integration range in Eq.~(\ref{eq:D3}) corresponds to the integration
along the direction of the diffracted wave, making an angle $\theta_{B}$
with the $x$-axis, from the bottom to the top surface of the crystal.
Since the integrand in Eq.~(\ref{eq:D3}) does not depend on $z$,
the integration along $s_{h}$ is replaced with the integration over
$x'$ by $ds_{h}=dx'/\cos\theta_{B}$.

The integral (\ref{eq:D3}) can also be written, by substituting $x'=x-z\cot\theta_{B}$,
as an integral over crystal thickness, 
\begin{equation}
\mathcal{E}_{h}^{\mathrm{kin}}(x)=i\frac{\pi\chi_{h}}{\lambda\sin\theta_{B}}\intop_{0}^{D}\exp\left[i\frac{Q(x-z\cot\theta_{B}){}^{2}}{2R}\right]\,dz.\label{eq:D6}
\end{equation}
Calculation of the integral is straightforward, 
\begin{equation}
\mathcal{E}_{h}^{\mathrm{kin}}(x)=i\frac{\pi\chi_{h}}{\lambda\cos\theta_{B}}s\left[F(x/s)-F((x-D\cot\theta_{B})/s)\right],\label{eq:D4}
\end{equation}
where $s^{2}=\pi|R|/Q$, and it is denoted 
\begin{equation}
F(x)=C(x)+i\sigma S(x).\label{eq:D5}
\end{equation}
Here $C(x)$ and $S(x)$ are cosine and sine Fresnel integrals, $\sigma=+1$
for $R>0$ (convex surface of bent crystal, as shown in Fig.~\ref{fig:geometry})
and $\sigma=-1$ for $R<0$ (concave crystal surface).

Green lines in Figs.~\ref{fig:dynkin}(a) and \ref{fig:dynkin}(c)
show kinematical intensity $\left|\mathcal{E}_{h}^{\mathrm{kin}}(x)\right|^{2}$
calculated with the same values of all parameters as in the corresponding
dynamical diffraction calculations. The kinematical intensity almost
coincides with the dynamical one, thus providing a final proof for
the applicability of the approximation of kinematical diffraction
for the curvature radii small compared with the critical radius. We
note that the coincidence of the curves is reached on the absolute
scale, without adjusting intensities.

\subsection{Fraunhofer diffraction}

The diffracted wave at the crystal surface $\mathcal{E}_{h}(x)$ transforms
during further propagation of the wave in free space to a detector.
At large enough distances from the diffracting crystal (Fraunhofer
diffraction), the x-ray wave field is described by the Fourier transform
of $\mathcal{E}_{h}(x)$. Let us consider the field distribution at
such distances, assuming that the field transformation in the $y$-direction
normal to the scattering plane is still not involved. Transformation
of the wave diffracted by the bent crystal on propagation in free
space over finite distances (Fresnel diffraction) is considered in
Sec.~\ref{sec:FresnelDiffraction}.

To obtain the Fourier spectrum of the kinematical diffraction amplitude
(\ref{eq:D3}), we represent it as a convolution integral 
\begin{equation}
\mathcal{E}_{h}^{\mathrm{kin}}(x)=i\frac{\pi\chi_{h}}{\lambda\cos\theta_{B}}\intop_{-\infty}^{\infty}\exp\left(i\frac{Qx'^{2}}{2R}\right)\Pi(x-x')\,dx',\label{eq:K14}
\end{equation}
where the function $\Pi(\xi)$ is defined as $\Pi=1$ for $0<\xi<D\cot\theta_{B}$
and $\Pi=0$ out of this interval. Making the Fourier transformation
of the two terms under the integral, we get 
\begin{equation}
\mathcal{E}_{h}^{\mathrm{kin}}(x)=\intop_{-\infty}^{\infty}\mathcal{E}^{\mathrm{Fraunhofer}}(q_{x})\exp(iq_{x}x)\,dq_{x},\label{eq:K15}
\end{equation}
where 
\begin{equation}
\mathcal{E}^{\mathrm{Fraunhofer}}(q_{x})=\mathrm{sinc}(q_{z}D/2)\exp\left(-i\frac{Rq_{x}^{2}}{2Q}-iq_{z}D/2\right),\label{eq:K16}
\end{equation}
$\mathrm{sinc}(x)=\sin(x)/x$, $q_{z}=q_{x}\cot\theta_{B}$, and a
constant prefactor is omitted in Eq.~(\ref{eq:K16}) to simplify
expressions. Intensity in the far field (Fraunhofer diffraction) is
given simply by $\mathrm{sinc}^{2}(q_{z}D/2)$, which provides a resolution
inversely proportional to the thickness $D$. Under conditions of
kinematical diffraction, it can be better than the resolution of dynamical
diffraction, which is limited by the extinction length. This resolution
is studied further in the next section.

\section{Spectral resolution}

\label{sec:SpectralResolution}

Equations in the previous section do not include an angular deviation
of the incident wave from the Bragg condition and are restricted with
the limit $\alpha=0$. To avoid these restrictions and also allow
a coherent superposition of waves with different wavelengths, we use
a more general expression for the kinematical diffraction amplitude
as an integral over the scattering plane of the crystal,

\begin{equation}
A(q_{x},q_{z})=\intop_{-\infty}^{\infty}\negthickspace dx\negthickspace\intop_{-D/2}^{D/2}\negthickspace dz\,\exp\left[-iq_{x}x-iq_{z}z+i\frac{Q(x^{2}+\alpha z^{2})}{2R}\right].\label{eq:K3}
\end{equation}
We restrict ourselves in this section to the Fraunhofer diffraction.
The wave vector $\mathbf{q}=\mathbf{K}^{\mathrm{out}}-\mathbf{K}^{\mathrm{in}}-\mathbf{Q}$
is the deviation of the scattering vector $\mathbf{K}^{\mathrm{out}}-\mathbf{K}^{\mathrm{in}}$
from the reciprocal lattice vector $\mathbf{Q}$. We have $\mathbf{q}=0$
for the wave of the reference wavelength incident on the crystal exactly
at the Bragg angle $\theta_{B}$ corresponding to that wavelength
and reflected at the Bragg angle. The components of the scattering
vector $\mathbf{q}=(q_{x},q_{z})$ depend on the angular deviations
$\delta\theta,\,\delta\theta'$of both incident and scattered waves,
and on the deviation $\delta k$ of the length of the wave vector
in the incident spectrum from the reference wave vector $k_{0}$ (since
scattering is elastic, the lengths of the wave vectors of the incident
and the scattered waves coincide). Explicit expressions for $q_{x}$
and $q_{z}$ are derived in Appendix~\ref{sec:AppendixB}. It is
convenient, for the purpose of comparison of the incident and the
diffracted spectra of an XFEL pulse, to represent the diffracted intensity
in an energy spectrum by considering the scattering angle $2\theta_{B}+\delta\theta+\delta\theta'$
as a Bragg angle for the respective wave vector $k_{0}+\delta k'$.
The components $q_{x},q_{z}$ of the scattering vector expressed through
the angular deviation of the incident beam $\delta\theta$ and the
wave vector deviations $\delta k,\delta k'$ are given by Eq.~(\ref{eq:K12}).
Particularly, an XFEL pulse can be described as a coherent superposition
of plane waves with different wavelengths propagating in the same
direction. With the crystal oriented at the Bragg angle for the reference
wavelength ($\delta\theta=0$), we get 
\begin{equation}
q_{x}=-2\delta k'\tan\theta_{B}\sin\theta_{B},\,\,\,q_{z}=2(\delta k'-\delta k)\sin\theta_{B}.\label{eq:K17}
\end{equation}

The $x$-dependent terms of the phase in the integral (\ref{eq:K3})
can be recollected as 
\begin{equation}
\frac{Qx^{2}}{2R}-q_{x}x=\frac{Q}{2R}(x-x_{0})^{2}-\frac{Rq_{x}^{2}}{2Q},\label{eq:K18}
\end{equation}
where $x_{0}=Rq_{x}/Q$. The exponential factor in the integral with
this phase strongly oscillates everywhere except an interval of the
width $\Delta x\sim(R/Q)^{1/2}$ around a point $x_{0}$. This range
of $x$ provides the main contribution to the integral. For a monochromatic
wave with an angular deviation $\delta\theta$ from the Bragg orientation,
we get from Eq.~(\ref{eq:K9}) that the center of the diffracting
region occurs at $x_{0}=R\delta\theta$. When the angular deviation
of the incident wave is so strong that $x_{0}$ exceeds the width
of the incident wave, the interval of $x$ contributing to diffraction
goes out of the illuminated part of the crystal, which causes a decrease
of the diffracted intensity and defines the rocking curve width of
the bent crystal. For smaller angular deviations, the interval $\Delta x$
is within the illuminated area, and the width $w$ does not restrict
diffraction. In Appendix~\ref{sec:AppendixC}, we explicitly calculate
the kinematical integral (\ref{eq:K3}) for a Gaussian profile of
the incident wave with a width $w$, as sketched in Fig.~\ref{fig:geometry}.
The resulting expression is rather bulky. In most cases of practical
interest, the width of the incident beam is so large that the outer
parts of the beam are out of diffraction, and the width $w$ does
not limit diffraction. We consider this latter case further on.

The range $\Delta x\sim(R/Q)^{1/2}$ of the diffracting region in
the bent crystal increases with the increasing curvature radius $R$.
However, the applicability of the kinematical approximation is limited
by the curvature radii $R\ll R_{c}$. Using the second expression
for $R_{c}$ in Eq.~(\ref{eq:D1}), we find that $\Delta x\ll(\Lambda/2)(\cot\theta_{B})^{1/2}$.
We do not consider very small Bragg angles and conclude that the range
of $x$ contributing to the integral (\ref{eq:K3}) is much smaller
than the extinction length $\Lambda$. This result provides an additional
insight into the origin of the kinematical diffraction in bent crystals:
as long as the condition (\ref{eq:D1}) of kinematical diffraction
is satisfied, the diffraction takes place in a narrow column of a
width $\Delta x\ll\Lambda$ in the crystal. The diffracted wave leaves
this column and cannot influence back on the transmitted wave, even
when the thickness exceeds the extinction length.

The kinematical integral (\ref{eq:K3}) splits into a product of two
integrals, one over $x$ and the other over $z$. Since we consider
the region $\Delta x$ to be within the illuminated area, the integral
over $x$ is calculated in infinite limits. The remaining integral
is over $z$, 
\begin{equation}
A(q_{x},q_{z})=\exp\left(-i\frac{Rq_{x}^{2}}{2Q}\right)\intop_{-D/2}^{D/2}\exp\left(-iq_{z}z+i\frac{\alpha Qz^{2}}{2R}\right)\,dz,\label{eq:K11}
\end{equation}
where we again omit a constant prefactor. When $\alpha=0$, Eq.~(\ref{eq:K11})
reduces to Eq.~(\ref{eq:K16}) but allows for more general expressions
(\ref{eq:K12}) for the components of the vector $\mathbf{q}$.

\begin{figure}
\includegraphics[width=\columnwidth]{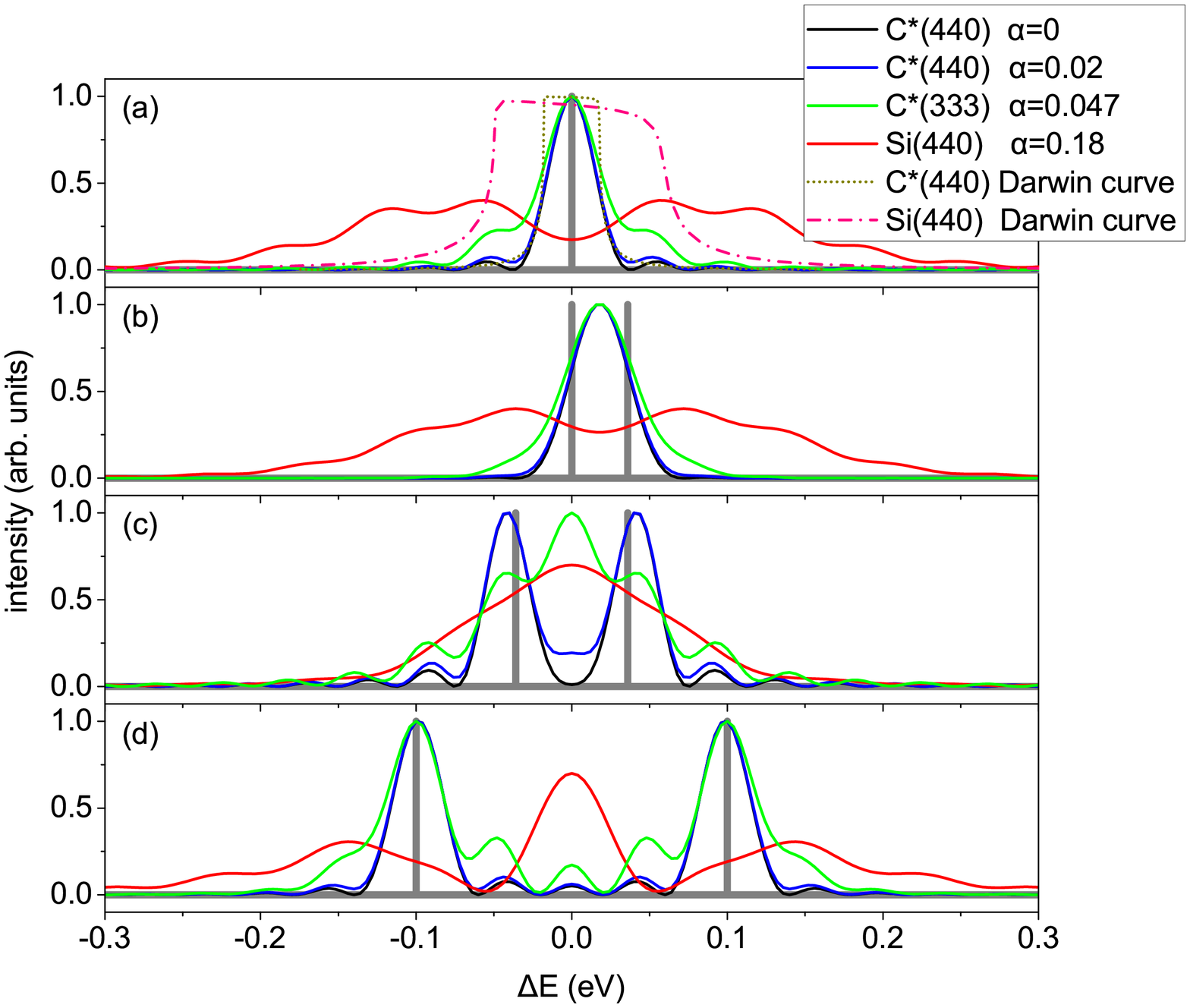}
\caption{Angular distributions of the waves diffracted by 20~\textmu m thick
crystal plates bent to a radius of 10~cm, calculated by Eq.~(\ref{eq:K19})
and represented in the same energy spectrum as the incident waves.
The incident spectra are shown by thick gray lines and consist of
(a) a single monochromatic plane wave of the energy 12~keV or (b)--(d)
two coherent monochromatic waves with small differences in wavelengths.
The dotted lines in (a) show the Darwin rocking curves for C{*}(440)
and Si(440) reflections.}
\label{fig:OneTwoPeaks} 
\end{figure}

Let us focus first on this limiting case $\alpha=0$, which is of
a special interest since it corresponds to the case of 110 oriented
diamond plate. In this case, the scattering intensity due to an incident
monochromatic plane wave is simply $\mathrm{sinc}^{2}(q_{z}D/2)$.
The intensity distribution is the same as in the classical problem
of diffraction grating in light optics. It is shown in Fig.~\ref{fig:OneTwoPeaks}(a)
by a black line.

The dotted line in Fig.~\ref{fig:OneTwoPeaks}(a) is the Darwin rocking
curve from a non-bent semi-infinite crystal in the same symmetric
Bragg reflection C{*}(440). Its full width at half-maximum is close
to that of a bent 20~\textmu m thick crystal. A thicker bent crystal
will provide a narrower curve. We note that its width does not depend
on the bending radius.

Figure \ref{fig:OneTwoPeaks}(a) also presents the angular distribution
of the waves diffracted from a 20~\textmu m thick silicon crystal, bent
to the same radius of 10~cm, in the same reflection 440, and the
Darwin curve for this reflection. Both curves are several times broader
than the respective curves in C{*}(440) reflection, but the reasons
of their broadening are different. A broader Darwin curve results
from a larger susceptibility and a smaller Bragg angle of the Si(440)
reflection with respect to the C{*}(440) reflection. The width of
the angular distribution of the waves diffracted by the bent crystal
does not depend on the susceptibility, because of the kinematical
diffraction, and the broader curve in Si(440) reflection is due to
a larger value of the parameter $\alpha$.

The possibility to resolve two waves with the same incidence direction
and different wavelengths is commonly defined in light optics by the
Rayleigh criterion (two wavelengths are resolved if the maximum of
diffracted intensity from one of them corresponds to the first minimum
of the other). In the case $\alpha=0$ this criterion, applied to
the sum of intensities $\mathrm{sinc}^{2}(q_{z}D/2)$ for two different
wavelengths, gives the resolution $\Delta q_{z}=2\pi/D$, and hence
$\Delta k=\pi/(D\sin\theta_{B})$. The resolution is limited by the
crystal thickness, which can be larger than the extinction length.
Hence, kinematical diffraction on a strongly bent crystal can provide
better resolution than the dynamical diffraction on a planar crystal.
The analysis above explains this surprising result: kinematical diffraction
on a strongly bent crystal takes place in a column whose width $\Delta x$
is small compared to the extinction length, but (for $\alpha=0$)
the height is equal to the crystal thickness $D$. That results in
a kinematical scattering from bent crystal whose thickness is not
limited by the extinction length. The energy resolution $\Delta E$
is related to the momentum resolution $\Delta k$ simply by $\Delta E/E=\Delta k/k$,
so that the Rayleigh criterion reads 
\begin{equation}
\Delta E/E=d/D,\label{eq:K20}
\end{equation}
where $d$ is the interplanar distance of the actual reflection and
the Bragg law $2d\sin\theta_{B}=\lambda$ is used. For our example
case, we get $\Delta E/E=3\times10^{-6}$ and $\Delta E=0.04$~eV.
The latter value is close to the width of the Darwin curve for a semi-infinite
non-bent crystal, see Fig.~ \ref{fig:OneTwoPeaks}(a).

\citeasnoun{boesenberg17} considered diffraction on a bent crystal
purely geometrically and arrived at a resolution defined by the pixel
size of a detector. Its contribution can be added to the diffraction
limited resolution (\ref{eq:K20}), when needed.

The energy spectrum of an XFEL pulse originates from the spectral
expansion of a short pulse, so that different wavelengths contribute
coherently and the amplitude $\mathcal{E}^{\mathrm{out}}(k')$ of
the electric field of the diffracted wave is related to the amplitude
$\mathcal{E}^{\mathrm{in}}(k)$ of the electric field incident on
the bent crystal by 
\begin{equation}
\mathcal{E}^{\mathrm{out}}(k')=\int A(k,k')\mathcal{E}^{\mathrm{in}}(k)dk,\label{eq:K21}
\end{equation}
where the diffraction amplitude $A(k,k')$ is described by equations
(\ref{eq:K19}) or (\ref{eq:K5}) with the components of the wave
vector $\mathbf{q}$ given by Eq.~(\ref{eq:K17}).

Figure \ref{fig:OneTwoPeaks}(b) presents angular distributions of
the waves diffracted by a bent crystal when the incident wave is a
coherent superposition of two monochromatic waves with the wavelength
difference corresponding to the Rayleigh criterion (\ref{eq:K20}).
The angular distributions are represented as corresponding spectra,
as described above. The two monochromatic components are not resolved
since the Rayleigh criterion is formulated for two incoherent waves
and implies the sum of intensities, rather than the sum of amplitudes.

Figure \ref{fig:OneTwoPeaks}(c) shows calculated angular distributions
of the diffracted waves for the wavelength difference between two
coherent monochromatic components two times larger than given by the
Rayleigh criterion. The components are well resolved. The resolution,
defined as the ability to resolve two monochromatic lines, in the
case of the coherent superposition of two waves occurs about 1.5 times
worse than given by the Rayleigh criterion (\ref{eq:K20}). Figure
\ref{fig:OneTwoPeaks}(d) shows calculated spectra for a larger wavelength
difference of the two monochromatic components of the incident wave.
The components are well resolved for $\alpha=0$ (black line).

The resolution (\ref{eq:K20}) is obtained by neglecting the second
term in the exponent in the integral (\ref{eq:K11}). This is possible
as long as $\alpha$ is so small that $\alpha Q(D/2)^{2}/2R$ is much
smaller than 1. In the general case $\alpha\neq0$, calculation of
the integral gives 
\begin{eqnarray}
A(q_{x},q_{z}) & = & \exp\left[-i\frac{R(q_{x}^{2}+\sigma q_{z}^{2}/\alpha)}{2Q}\right]\label{eq:K19}\\
 &  & \times\left[F\left(\frac{q_{z}+aD}{\sqrt{2\pi a}}\right)-F\left(\frac{q_{z}-aD}{\sqrt{2\pi a}}\right)\right],\nonumber 
\end{eqnarray}
where $a=\alpha Q/2|R|$ and the function $F(x)$ is defined in Eq.~(\ref{eq:D5}).

For C{*}(440) reflection, the factor $\alpha Q(D/2)^{2}/2R$ is approximately
equal to 1, and the calculated curves (blue curves in Fig.~\ref{fig:OneTwoPeaks})
are close to the calculation with $\alpha=0$ (black curves). This
factor calculated for C{*}(333) reflection (with $\alpha=0.047)$
is approximately equal to 2, which already results in a notable modification
of the diffraction curves (green curves in Fig.~\ref{fig:OneTwoPeaks}).
For Si(440) reflection with $\alpha=0.18$, this factor is 5.9, which
results in complicated diffraction patterns (red curves in Fig.~\ref{fig:OneTwoPeaks}),
rather than a broadening of the corresponding spectral lines.

Figure \ref{fig:OneTwoPeaks} shows that, due to a coherent superposition
of the monochromatic components, the worse resolution for $\alpha\neq0$
cannot be described as the broadening of the sharp peaks of the incident
spectrum. Rather, a complicated interference pattern arises, and the
incident spectrum can hardly be recognized in it. The width of the
interference fringes is still given by Eq.~(\ref{eq:K20}).

\section{Fresnel diffraction}

\label{sec:FresnelDiffraction}

In this section, we consider the finite-distance free-space propagation
of the wave diffracted by a bent crystal. This allows us to establish
the applicability limits of the Fraunhofer approximation used in the
previous section and evaluate corrections due to a finite distance
from the bent crystal to a detector.

Let us follow the free space propagation of the electric field at
the crystal surface $\mathcal{E}_{h}^{\mathrm{kin}}(x)$ given by
Eqs.~(\ref{eq:D3})--(\ref{eq:D4}) for the case $\alpha=0$. At
a distance $L$ from the bent crystal, the free space propagation
is described {[}see e.g. \citeasnoun{born+wolf}, \S8.3, and \citeasnoun{cowley75book},
\S1.7{]} by multiplying the electric field at the crystal surface
$\mathcal{E}_{h}^{\mathrm{kin}}(x)$ with the phase factor $\exp(i\pi\xi^{2}/\lambda L)$,
where $\xi$ is the distance in the direction perpendicular to the
propagation direction of the diffracted beam, $\xi=x\sin\theta_{B}$:
\begin{equation}
\mathcal{E}^{\mathrm{Fresnel}}(q_{x})=\intop_{-\infty}^{\infty}\mathcal{E}_{h}^{\mathrm{kin}}(x)\exp\left[i\frac{\pi\left(x\sin\theta_{B}\right)^{2}}{\lambda L}-iq_{x}x\right]\,dx.\label{eq:K22}
\end{equation}
Substituting here Eq\@.~(\ref{eq:D6}) and performing integration
over $x$, we represent Eq\@.~(\ref{eq:K22}) as 
\begin{equation}
\mathcal{E}^{\mathrm{Fresnel}}(q_{x})=\exp\left(-i\frac{\tilde{R}q_{x}^{2}}{2Q}\right)\intop_{-D/2}^{D/2}\exp\left(-i\tilde{q}_{z}z+i\frac{\tilde{\alpha}Qz^{2}}{2R}\right)\,dz,\label{eq:K23}
\end{equation}
where it is defined 
\begin{equation}
\tilde{R}=R\left(1+\frac{R\sin\theta_{B}}{2L}\right)^{-1}.\label{eq:K24}
\end{equation}
The quantities $\tilde{q}_{z}=(\tilde{R}/R)q_{x}\cot\theta_{B}$ and
$\tilde{\alpha}=\tilde{R}\cos^{2}\theta_{B}/(2L\sin\theta_{B})$ are
introduced here for the particular case $\alpha=0$. Below in Eq.~(\ref{eq:F3})
they are derived for the general case $\alpha\neq0$. In the limit
$L\rightarrow\infty$, the Fresnel diffraction amplitude (\ref{eq:K23})
reduces to the Fraunhofer one (\ref{eq:K16}).

The distance $L$ required to reach the Fraunhofer limit follows from
Eqs.~(\ref{eq:K23}) and (\ref{eq:K24}). The first requirement is
$L\gg R\sin\theta_{B}$, which gives $\tilde{R}\approx R$. Since
Eq.~(\ref{eq:K23}) is written for $\alpha=0$, the second requirement
follows from the possibility to neglect the second term in the exponent
in the integral (\ref{eq:K23}). This term at $z=D/2$ is equal to
$\frac{\pi}{4}(D\cos\theta_{B})^{2}/\lambda L$. We note that the
crystal thickness $D$ seen from the direction of the diffracted beam
is $D\cos\theta_{B}$, while the diameter of the first Fresnel zone
is $\sqrt{\lambda L}$. Hence, the crystal thickness seen from the
direction of the diffracted beam should be smaller than the diameter
of the first Fresnel zone, i.e. the distances from crystal to detector
should be $L>(D\cos\theta_{B})^{2}/\lambda$. The minimum distance
depends on the Bragg angle: for our reference case of C{*}(440) reflection
at 12~keV and crystal thickness $D=20$~\textmu m, we get $L>1.3$~m,
while, for C{*}(220) at the same conditions, we have $L>3.2$~m. The
points in Fig.~\ref{fig:Rc} mark, for each reflection, the energy
given by the condition $\sqrt{\lambda L}=D\cos\theta_{B}$ for the
crystal thickness $D=20$~\textmu m and distance to detector $L=1$~m.
For energies smaller than marked, Fraunhofer approximation is approached
at 1~m distance to detector. Larger energies correspond to Fresnel
diffraction at such a distance.

Calculation of the integral (\ref{eq:K23}) gives

\begin{eqnarray}
\mathcal{E}^{\mathrm{Fresnel}}(q_{x}) & = & \exp\left[-i\frac{\tilde{R}q_{x}^{2}+\tilde{\sigma}R\tilde{q}_{z}^{2}/\tilde{\alpha}}{2Q}\right]\label{eq:F4}\\
 &  & \times\left[F\left(\frac{\tilde{q}_{z}+\tilde{a}D}{\sqrt{2\pi\tilde{a}}}\right)-F\left(\frac{\tilde{q}_{z}-\tilde{a}D}{\sqrt{2\pi\tilde{a}}}\right)\right].\nonumber 
\end{eqnarray}
Since the bending radius $R$ can be positive (convex surface of bent
crystal) or negative (concave crystal surface), we define a positive
quantity $\tilde{a}=|\tilde{\alpha}|Q/2|R|$ and the sign term $\tilde{\sigma}=+1$
if $\tilde{\alpha}$ and $R$ are of the same sign and $\tilde{\sigma}=-1$
if $\tilde{\alpha}$ and $R$ have opposite signs. The function $F(x)=C(x)+i\tilde{\sigma}S(x)$
is defined similarly to Eq.~(\ref{eq:D5}).

\begin{figure}
\includegraphics[width=\columnwidth]{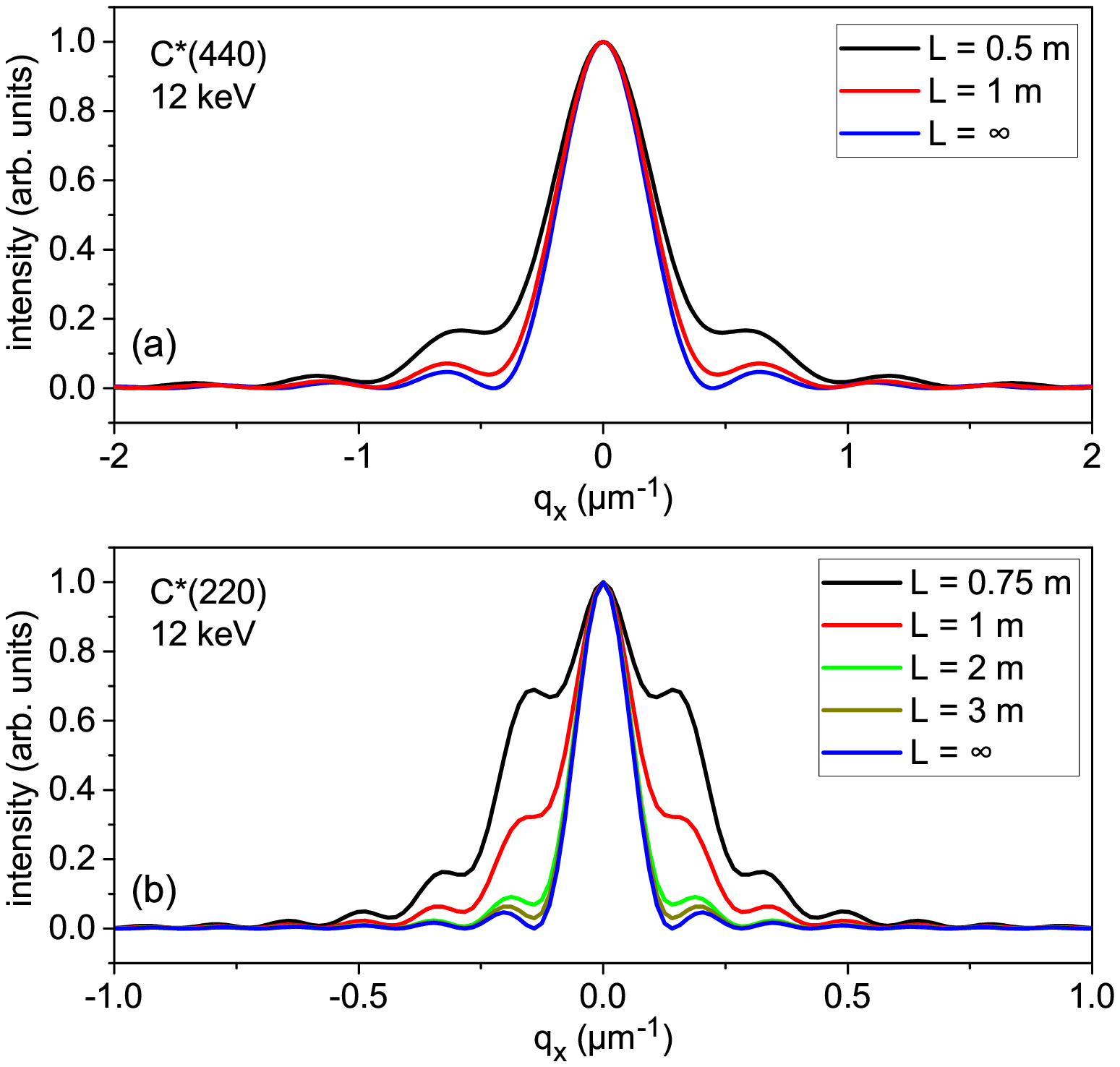}
\caption{Transformation on the way to detector of a monochromatic wave diffracted
from a 20~\textmu m thick diamond crystal bent to a radius of 10~cm,
calculated by Eq.~(\ref{eq:F4}). Reflections (a) 440 and (b) 220
are compared.}
\label{fig:FresnelFraunhofer} 
\end{figure}

Figure \ref{fig:FresnelFraunhofer} shows transformation of the diffracted
beam with the distance to detector, calculated by Eq.~(\ref{eq:F4}).
Reflections 440 and 220 from diamond at the same energy 12~keV are
compared. The only essential difference between reflections is their
Bragg angles: the larger Bragg angle of the 440 reflection gives rise
to smaller distances needed to reach the Fraunhofer diffraction range.

In the analysis above, we used the amplitude of the wave diffracted
by bent crystal $\mathcal{E}_{h}^{\mathrm{kin}}(x)$ that was written
for a monochromatic incident wave, exact Bragg orientation of the
incident wave, and the special case $\alpha=0$. In the general case
of the kinematical scattering amplitude (\ref{eq:K3}), the free space
propagation is described by an additional phase term $\exp(i\pi\xi^{2}/\lambda L),$where
$\xi=x\sin\theta_{B}+z\cos\theta_{B}$ is the distance in the direction
perpendicular to the beam diffracted by the crystal. Then, the amplitude
of the diffracted wave at the detector is written as 
\begin{eqnarray}
A(q_{x},q_{z}) & = & \intop_{-\infty}^{\infty}\negthickspace dx\negthickspace\intop_{-D/2}^{D/2}\negthickspace dz\,\exp\left[-iq_{x}x-iq_{z}z+i\frac{Q(x^{2}+\alpha z^{2})}{2R}\right]\nonumber \\
 &  & \times\exp\left[i\frac{\pi(x\sin\theta_{B}+z\cos\theta_{B})^{2}}{\lambda L}\right],\label{eq:F1}
\end{eqnarray}
which replaces the respective integral (\ref{eq:K3}) written for
Fraunhofer diffraction. The integral (\ref{eq:F1}) can be written
in the same form as Eq.~(\ref{eq:K23}) with the same expression
for $\tilde{R}$ given by Eq.~(\ref{eq:K24}) but $\tilde{q}_{z}$
and $\tilde{\alpha}$ are generalized as follows: 
\begin{eqnarray}
\tilde{q}_{z} & = & q_{z}-\frac{\tilde{R}\cos\theta_{B}}{2L}q_{x},\nonumber \\
\tilde{\alpha} & = & \alpha+\frac{\tilde{R}\cos^{2}\theta_{B}}{2L\sin\theta_{B}}.\label{eq:F3}
\end{eqnarray}
Calculation of the integral gives rise to Eq.~(\ref{eq:F4}). It
has the same form as the Fraunhofer amplitude (\ref{eq:K19}) but
with the parameters modified according to Eqs.~(\ref{eq:K24}) and
(\ref{eq:F3}).

We have already seen in the analysis of Fraunhofer diffraction in
Sec.~\ref{sec:SpectralResolution}, and in particular in Fig.~\ref{fig:OneTwoPeaks},
that the value of parameter $\alpha$ plays an essential role in spectral
resolution. Finite-distance free-space propagation of the wave diffracted
from the bent crystal gives rise to a modification of this parameter
to $\tilde{\alpha}$, as given by Eq.~(\ref{eq:F3}). In particular,
the concave bending ($R<0$) and appropriately chosen distance $L$
can be used to reduce this parameter and hence improve the resolution.

\begin{figure}
\includegraphics[width=\columnwidth]{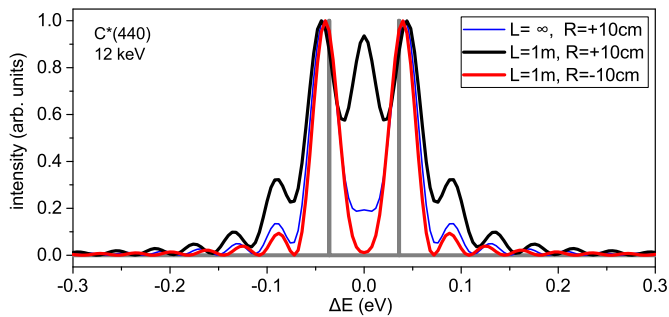}
\caption{Spectra of diffracted waves for an incident wave consisting of two
mutually coherent plane waves of different wavelengths (shown by thick
gray lines) for an infinite distance to detector (Fraunhofer diffraction,
blue line) and the distance to detector $L=1$~m (Fresnel diffraction),
calculated by Eq.~(\ref{eq:F4}). Symmetric Bragg reflection 440
from a 20~\textmu m thick diamond plate, bending radius 10~cm, convex
(black line) and concave (red line) bending are compared.}
\label{fig:Fresnel} 
\end{figure}

Figure \ref{fig:Fresnel} shows calculated spectra of diffracted waves
for an incident wave consisting of two coherent plane waves, the same
as in Fig.~\ref{fig:OneTwoPeaks}(c). The blue line in Fig.~\ref{fig:Fresnel}
is calculated for an infinite distance $L$ and represents the same
line in Fig.~\ref{fig:OneTwoPeaks}(c). Black and red lines are calculated
for a distance from bent crystal to detector of $L=1$~m. Calculation
by Eq.~(\ref{eq:F3}) gives $\tilde{\alpha}=0.039$ for a convex
bending with $R=+10$~cm and $\tilde{\alpha}=-0.00095$ for a concave
bending with $R=-10$~cm. The increase of $\tilde{\alpha}$ for the
convex bending has the same effect as an increase in $\alpha$ for
reflection C{*}(333) in Fig.~\ref{fig:OneTwoPeaks} and gives rise
to a more complicated spectrum with several fringes. The decrease
of $\tilde{\alpha}$ for the concave bending has an opposite effect
and leads to a simple spectrum of two waves described by Eq.~(\ref{eq:K16})
with the resolution given by Eq.~(\ref{eq:K20}).

\section{Spectra of XFEL pulses}

The spectra in the self-amplified spontaneous emission (SASE) mode
of the European XFEL have been generated with the simulation code
FAST \cite{saldin99}, which provides a 2D distribution of electric
field in real space at the exit of the undulator for each moment of
time for various parameters of the electron bunch charge and the undulator.
Simulation results are stored in an in-house database \cite{XFEL:FAST}.
The spectra are simulated for the electron energy 14~GeV, photon
energy 12.4~keV, and the active undulator length corresponding to
the saturation length, the point with the maximum brightness, for
a given electron bunch charge \cite{exfel-fel2014}.

Conversion from the time to the frequency domain has been performed
using the WavePropaGator package \cite{samoylova16}, which provides
a 2D distribution of electric field for each frequency of the pulse.
We use the spectrum at the center of the pulse in frequency domain,
assuming this distribution to be the same across the beam.

\begin{figure}
\includegraphics[width=\textwidth]{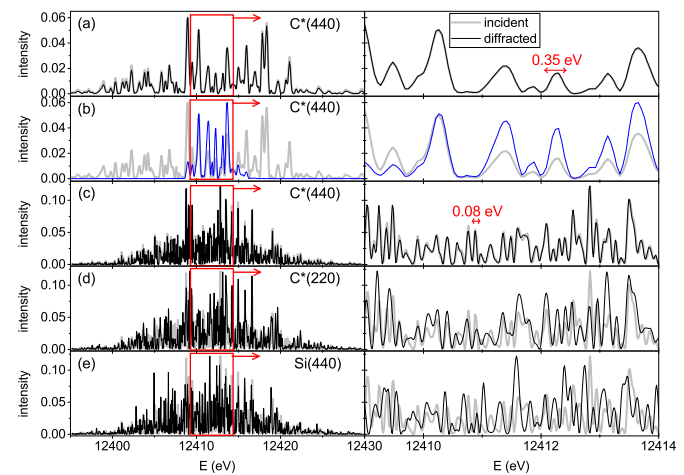}
\caption{Spectra of the waves incident on (a--d) diamond or (e) silicon plate
of thickness $D=20$~\textmu m bent to a radius of $R=10$~cm (thick
gray lines) and spectra of the diffracted waves in Fraunhofer diffraction
case (thin black or blue lines). The incident beam width is $w=500$~\textmu m
(a,c--e) or 50 \textmu m (b). The pulse duration is 10~fs (a,b)
and the undulator length is 75~m, or the pulse duration is 42~fs
(c--e) and the undulator length is 105~m. Spectrum of the incident
wave is convoluted, according to Eq.~(\ref{eq:K21}), with the scattering
amplitude given by equations (\ref{eq:K19}) for (a,c-e) or (\ref{eq:K5})
for (b).}
\label{fig:spectra} 
\end{figure}

Figure \ref{fig:spectra} compares spectra of the XFEL pulses incident
on the diffracting bent crystal (thick gray lines) and the spectra
of the diffracted waves (thin black or blue lines). Complex amplitudes
of the incident beams were used in calculation of diffraction by Eq.~(\ref{eq:K21}),
squared moduli of the amplitudes are shown in the figure and the respective
phases are not shown. Calculations of the diffraction amplitude $A(k,k')$
using Eq.~(\ref{eq:K19}) for an infinite width of the incident wave
or using Eq.~(\ref{eq:K5}) taking into account the finite width
of the incident beam give identical results for the width $w=500$~\textmu m
in Fig.~\ref{fig:spectra}(a),(c)--(e). For the width $w=50$~\textmu m
of a focused beam in Fig.~\ref{fig:spectra}(b), the equation (\ref{eq:K5})
is used. The bending radius of the crystal is taken $R=10$~cm and
its thickness $D=20$~\textmu m.

Figures \ref{fig:spectra}(a,b) show by thick gray lines a spectrum
of the XFEL pulse of the duration of approximately 10~fs generated
in an undulator of the active length 75~m. The pulse duration of
10~fs gives rise to a 0.35~eV characteristic width of the oscillations
in the spectrum. The numbers above are the full width at half maxima
(FWHM) of the peaks in time and frequency domains, respectively. Such
a spectrum is well resolved by the bent-crystal spectrometer in the
C{*}(440) reflection, as shown in Figs.~\ref{fig:spectra}(a,b).
A width of $w=500$~\textmu m of the incident beam is needed to resolve
the whole spectrum, see Fig.~\ref{fig:spectra}(a). If the beam is
focused to a width $w=50$~\textmu m, only a small part of the spectrum
is diffracted, see Fig.~\ref{fig:spectra}(b). The characteristic
width of the oscillations in the spectrum is still reproduced, and
hence the pulse duration can be estimated.

Figures \ref{fig:spectra}(c)--(e) show a spectrum of the x-ray pulse
of duration 42~fs at the undulator length 105~m. This pulse duration
gives rise to a 0.08~eV characteristic width of the oscillations
in the spectrum. The resolution of the bent-crystal spectrometer,
estimated with the Rayleigh criterion (\ref{eq:K20}), is about 0.04~eV.
The continuous spectrum of the x-ray pulse is fully reproduced in
C{*}(440) reflection, see Fig.~\ref{fig:spectra}(c). The reflection
C{*}(220), shown in Fig.~\ref{fig:spectra}(d), possesses, as it
follows from Eq.~(\ref{eq:K20}), two times worse resolution because
of the two times larger interplanar distance $d$. The initial spectrum
is not reproduced and its oscillations are not fully resolved. However,
the oscillations are of almost the same width as in the initial spectrum.
They can be used to estimate the pulse duration in time domain with
almost the same accuracy as the initial spectrum. In reflection Si(440)
presented in Fig.~\ref{fig:spectra}(e), the depth dependence of
the displacement field due to the value of $\alpha=0.18$ for silicon
gives rise to a worse resolution. The initial spectrum is not reproduced
but, as in the case of C{*}(220) reflection, the oscillations can
be used to estimate the pulse duration.

\begin{figure}
\includegraphics[width=\columnwidth]{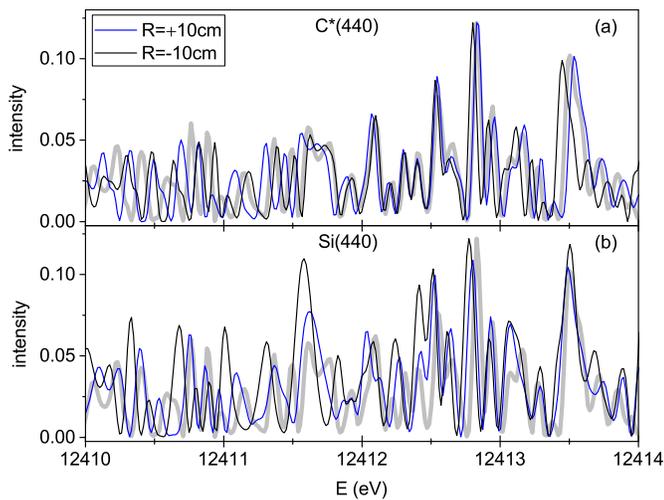}
\caption{The incident (thick gray lines) and diffracted (thin black and blue
lines) spectra at a distance $L=1$~m from a 20~\textmu m thick
diamond (a) or silicon (b) plate bent to a radius $R=10$~cm. Spectrum
of the incident wave is convoluted, according to Eq.~(\ref{eq:K21}),
with the scattering amplitude given by Eq.~(\ref{eq:F4}).}
\label{fig:spectraFresnel} 
\end{figure}

Figures \ref{fig:spectraFresnel} compares spectra calculated for
a distance $L=1$~m from the bent crystal to a detector, for C{*}(440)
and Si(440) reflections for the same incident pulse as in Figs.~\ref{fig:spectra}(c-e).
Bending in opposite directions, concave and convex, are compared for
each reflection. For C{*}(440) reflection, the spectrum is somewhat
expanded (at $R>0$) or compressed (at $R<0$) with respect to the
spectrum of the incident pulse. For Si(440) reflection, transformation
of the spectrum is more complicated, but it does not change the structure
of the spectrum qualitatively.

In all cases presented in Figs.~\ref{fig:spectra} and \ref{fig:spectraFresnel},
the spectra of the waves diffracted from a bent crystal are qualitatively
similar to the spectra of the incident beams. The widths of the fringes
in the spectra can be used to estimate duration of the incident pulses.
However, only C{*}(440) reflection reproduces the incident spectrum
at the energy of 12 keV. Even in this case, the spectrum is slightly
expanded or compressed, depending on the direction of bending, due
to a finite distance from bent crystal to detector.

For other reflections, spectra of the waves diffracted by the bent
crystal do not coincide with the Fourier transformations of the incident
pulses. However, when the conditions for kinematical diffraction are
satisfied, they can be calculated for a given incident pulse using
diffraction amplitudes derived above and used in a fitting procedure
to obtain time structure of the incident pulse.

\section{Conclusions}

X-ray diffraction from a bent single crystal can be treated kinematically
when the bending radius is small compared to the critical radius given
by the ratio of the Bragg-case extinction length for the actual reflection
to the Darwin width of this reflection. The critical radius varies,
depending on the x-ray energy, the crystal, and the reflection chosen,
from centimeters to meters.

Under conditions of kinematical diffraction, each monochromatic component
of the pulse finds diffraction conditions only in a column inside
the crystal with the width much smaller than the extinction length.
In a cylindrically bent diamond plate of 110 orientation, the entire
column diffracts in phase, since the Poisson effect on bending is
compensated by the elastic anisotropy, and the displacement field
does not vary over the depth. In this case, the spectral resolution
is limited by the crystal thickness, rather than the extinction length,
and can be better than the resolution of a planar dynamically diffracting
crystal. It amounts to the ratio of the lattice spacing for the actual
reflection to the crystal thickness. As an example, the symmetric
Bragg reflection 440 from diamond provides almost undistorted spectrum
for x-ray energies of about 12~keV with the resolution of 0.04~eV.

The spectrum of the waves diffracted by the bent crystal generally
differs from the spectrum of the incident pulse. Hence, the spectrum
is not resolved in a rigorous spectroscopic sense. However, the diffracted
spectra look qualitatively similar to the respective incident spectra.
The widths of their fringes can still be used to estimate duration
of the incident x-ray pulse. A finite distance from the bent crystal
to a detector (Fresnel diffraction) causes additional modifications
of the measured spectrum, but still leaves it qualitatively similar
to the incident one.

\ack{The authors thank Alexander Belov, Andrei Benediktovitch, Ulrike
Boesenberg, Anders Madsen, Harald Sinn, Sergey Terentyev, and Thomas
Tschentscher for useful discussions, and Bernd Jenichen and Kurt Ament
for critical reading of the manuscript.}

\appendix

\section{Displacement field in a bent anisotropic thin plate}

\label{sec:AppendixA}

\subsection{Elastic equilibrium equations and their solution}

To calculate the displacement field in a bent plate, taking into account
its elastic anisotropy, we begin with the Hooke's law in the $6\times6$
formulation $\epsilon_{m}=s'_{mn}\sigma_{n}$, where $m,n$ denote
pairs of indices ($1\rightarrow11,\,2\rightarrow22,\,3\rightarrow33,\,4\rightarrow23,\,5\rightarrow13,\,6\rightarrow12$).
Here, $s'_{mn}$ are the components of the compliance tensor, and
the prime denotes the components in the coordinate system with the
$x,y$ axes in the plane of the plate and the $z$ axis normal to
it. The notation $s_{mn}$ without the prime is reserved for the components
of the compliance tensor in the standard cubic reference frame. The
components of the stress tensor are denoted by $\sigma_{n}$, and
the components of the strain tensor $\epsilon_{n}$ are written in
the engineering notation (i.e., without the coefficient 1/2 at the
off-diagonal components): 
\begin{align}
\epsilon_{1}=\frac{\partial u_{x}}{\partial x}, & \,\,\,\,\,\,\,\epsilon_{4}=\frac{\partial u_{y}}{\partial z}+\frac{\partial u_{z}}{\partial y},\nonumber \\
\epsilon_{2}=\frac{\partial u_{y}}{\partial y}, & \,\,\,\,\,\,\,\epsilon_{5}=\frac{\partial u_{x}}{\partial z}+\frac{\partial u_{z}}{\partial x},\label{eq:A1}\\
\epsilon_{3}=\frac{\partial u_{z}}{\partial z}, & \,\,\,\,\,\,\,\epsilon_{6}=\frac{\partial u_{x}}{\partial y}+\frac{\partial u_{y}}{\partial x}.\nonumber 
\end{align}
The absence of forces at the plate surface gives $\sigma_{iz}=0$
(where $i=1,2,3$) and, since the plate is thin, these components
of stress are small in comparison with the other stress components
also inside the plate, so that $\sigma_{3}=\sigma_{4}=\sigma_{5}=0$.

We consider bending of the plate by two moments, $M_{1}$ about the
$y$ axis and $M_{2}$ about the $x$ axis, which give rise to stress
linearly varying across the plate {[}see \cite{lekhnitskii81}, Eq.~(16.1){]}:
\begin{equation}
\sigma_{1}=\frac{12M_{1}}{D^{3}}z,\,\,\,\,\,\,\sigma_{2}=\frac{12M_{2}}{D^{3}}z,\label{eq:A2}
\end{equation}
where $D$ is the plate thickness. We do not include torsion in the
consideration and hence take $\sigma_6=0$. Thus, the components
$\sigma_{1}$ and $\sigma_{2}$ in Eq.~(\ref{eq:A2}) are the only
nonzero stress components, and the elastic equilibrium equations read
\begin{align}
\frac{\partial u_{x}}{\partial x}=s'_{11}\sigma_{1}+s'_{12}\sigma_{2}, & \,\,\,\,\,\,\,\frac{\partial u_{y}}{\partial z}+\frac{\partial u_{z}}{\partial y}=s'_{14}\sigma_{1}+s'_{24}\sigma_{2},\nonumber \\
\frac{\partial u_{y}}{\partial y}=s'_{12}\sigma_{1}+s'_{22}\sigma_{2}, & \,\,\,\,\,\,\,\frac{\partial u_{x}}{\partial z}+\frac{\partial u_{z}}{\partial x}=s'_{15}\sigma_{1}+s'_{25}\sigma_{2},\nonumber \\
\frac{\partial u_{z}}{\partial z}=s'_{13}\sigma_{1}+s'_{23}\sigma_{2}, & \,\,\,\,\,\,\,\frac{\partial u_{x}}{\partial y}+\frac{\partial u_{y}}{\partial x}=s'_{16}\sigma_{1}+s'_{26}\sigma_{2}.\label{eq:A3}
\end{align}
The solution of these equations, with the center of the plate fixed
at zero ($u_{x}=u_{y}=u_{z}=0$ at the point $x=y=z=0$) is (see \citeasnoun{lekhnitskii81},
Eq.~(16.3)): 
\begin{eqnarray}
u_{x} & = & \frac{6}{D^{3}}[M_{1}(s'_{15}z^{2}+s'_{16}yz+2s'_{11}xz)\nonumber \\
 &  & +M_{2}(s'_{25}z^{2}+s'_{26}yz+2s'_{12}xz)],\nonumber \\
u_{y} & = & \frac{6}{D^{3}}[M_{1}(s'_{14}z^{2}+2s'_{12}yz+s'_{16}xz)\nonumber \\
 &  & +M_{2}(s'_{24}z^{2}+2s'_{22}yz+s'_{26}xz)],\nonumber \\
u_{z} & = & \frac{6}{D^{3}}[M_{1}(s'_{13}z^{2}-s'_{11}x^{2}-s'_{12}y^{2}-s'_{16}xy)\nonumber \\
 &  & +M_{2}(s'_{23}z^{2}-s'_{12}x^{2}-s'_{22}y^{2}-s'_{26}xy)].\label{eq:A4}
\end{eqnarray}

For the symmetric Bragg case diffraction considered in the present
work, only the displacement normal to the plate plane $u_{z}$ is
of interest. Crystal orientations that we consider (see the next section)
give $s'_{16}=s'_{26}=0$. Then, the displacement for a biaxial bending
can be written in the form $u_{z}=x^{2}/2R_{x}+y^{2}/2R_{y}+Kz^{2}$.
The curvature radii $R_{x}$ and $R_{y}$ are easily derived from
Eq.~(\ref{eq:A4}), and we do not present these bulk expressions
here. Rather, we consider the case of cylindrical bending, when the
bending moments $M_{1}$ and $M_{2}$ are applied to provide $R_{y}\rightarrow\infty$.
In this case, the displacement field can be written as 
\begin{equation}
u_{z}=(x^{2}+\alpha z^{2})/2R,\label{eq:A8}
\end{equation}
where 
\begin{equation}
\alpha=\frac{s'_{12}s'_{23}-s'_{13}s'_{22}}{s'_{11}s'_{22}-(s'_{12})^{2}}.\label{eq:A9}
\end{equation}
We omit a bulky expression of the curvature radius $R$ through the
bending moments, since the radius, rather than the moments, is directly
measured in the experiment.

The aim of the next sections is to calculate coefficient $\alpha$
for different crystallographic orientations of the plate, and for
different materials. For that purpose, the compliances $s'_{mn}$
need to be calculated for the respective crystallographic orientations.

\subsection{Transformation of the compliance tensor}

Transformation of the components $s_{mn}$ from the reference coordinate
system with the standard axes of the cubic crystal to the coordinate
system related to the crystallographic orientation of the plate requires
rotation of the 4th-rank tensor $s_{ijkl}$ to the new coordinate
system by four rotation matrices. \citeasnoun{wortman65} and \citeasnoun{lekhnitskii81}(\S5)
proposed two different practical methods to make this transformation.
We did not make a thorough check of the equivalence of these methods
but applied both of them to the orientations that are of interest
for us and ascertained that they give identical results in these cases.

The elastic compliances tensor for a cubic crystal in the standard
reference frame with $\left\langle 100\right\rangle $ axes is

\begin{equation}
s=\begin{pmatrix}s_{11} & s_{12} & s_{12} & 0 & 0 & 0\\
s_{12} & s_{11} & s_{12} & 0 & 0 & 0\\
s_{12} & s_{12} & s_{11} & 0 & 0 & 0\\
0 & 0 & 0 & s_{44} & 0 & 0\\
0 & 0 & 0 & 0 & s_{44} & 0\\
0 & 0 & 0 & 0 & 0 & s_{44}
\end{pmatrix}.\label{eq:A5}
\end{equation}

For 110 oriented plate, namely, $x$ axis along $[1\bar{1}0],$ $y$
axis along $[001]$, and $z$ axis along $[110]$, we obtain 
\begin{equation}
s'=\begin{pmatrix}s_{11}-s_{c}/2 & s_{12} & s_{12}+s_{c}/2 & 0 & 0 & 0\\
s_{12} & s_{11} & s_{12} & 0 & 0 & 0\\
s_{12}+s_{c}/2 & s_{12} & s_{11}-s_{c}/2 & 0 & 0 & 0\\
0 & 0 & 0 & s_{44} & 0 & 0\\
0 & 0 & 0 & 0 & s_{44}+2s_{c} & 0\\
0 & 0 & 0 & 0 & 0 & s_{44}
\end{pmatrix},\label{eq:A6}
\end{equation}
where it is defined 
\begin{equation}
s_{c}=s_{11}-s_{12}-s_{44}/2.\label{eq:A11}
\end{equation}

For 111 oriented plate, namely $x$ axis along $[1\bar{1}0]$, $y$
axis along $[11\bar{2}]$, and $z$ axis along $[111]$, we find \onecolumn
\begin{equation}
s'=\begin{pmatrix}s_{11}-s_{c}/2 & s_{12}+s_{c}/6 & s_{12}+s_{c}/3 & \sqrt{2}s_{c}/3 & 0 & 0\\
s_{12}+s_{c}/6 & s_{11}-s_{c}/2 & s_{12}+s_{c}/3 & -\sqrt{2}s_{c}/3 & 0 & 0\\
s_{12}+s_{c}/3 & s_{12}+s_{c}/3 & s_{11}-2s_{c}/3 & 0 & 0 & 0\\
\sqrt{2}s_{c}/3 & -\sqrt{2}s_{c}/3 & 0 & s_{44}+4s_{c}/3 & 0 & 0\\
0 & 0 & 0 & 0 & s_{44}+4s_{c}/3 & 2\sqrt{2}s_{c}/3\\
0 & 0 & 0 & 0 & 2\sqrt{2}s_{c}/3 & s_{44}+2s_{c}/3
\end{pmatrix}.\label{eq:A7}
\end{equation}
\twocolumn

\subsection{Bending of diamond and silicon plates}

Below we use the literature values of the elastic moduli $c_{mn}$
and calculate the compliances $s_{mn}$ for the $\left\langle 100\right\rangle $
reference frame as 
\begin{eqnarray}
s_{11} & = & \frac{c_{11}+c_{12}}{c_{11}^{2}+c_{11}c_{12}-2c_{12}^{2}},\nonumber \\
s_{12} & = & -\frac{c_{12}}{c_{11}^{2}+c_{11}c_{12}-2c_{12}^{2}},\nonumber \\
s_{44} & = & \frac{1}{c_{44}}.\label{eq:A10}
\end{eqnarray}

We consider now two materials, diamond and silicon, which are used
in spectrometers for XFELs. The elastic moduli of diamond are \cite{mcskimin72}
$c_{11}=10.79$, $c_{12}=1.24$, $c_{44}=5.78$ and of silicon \cite{wortman65}
$c_{11}=1.657$, $c_{12}=0.639$, $c_{44}=0.796$ (all in units $10^{12}$~dyn/cm$^{2}$).

Using the compliances (\ref{eq:A6}) for 110 oriented plate, we obtain
the coefficient $\alpha$ in Eq.~(\ref{eq:A9}) equal to $\alpha=0.020$
for diamond and $\alpha=0.18$ for silicon. The calculation for 111
oriented plane using Eq.~(\ref{eq:A7}) gives $\alpha=0.047$ for
diamond and $\alpha=0.22$ for silicon. Thus, the elastic properties
of diamond give rise to an exceptionally small variation of strain
over the depth $z$.

To understand the origin of the small coefficient $\alpha$ for 110
oriented diamond plate, we express it through the Poisson ratio $\nu=-s_{12}/s_{11}$
and the Zener anisotropy ratio $A=2(s_{11}-s_{12})/s_{44}$. Then,
the coefficient $\alpha$ in Eq.~(\ref{eq:A9}) for 110 oriented
plate can identically be written as 
\begin{equation}
\alpha=\frac{\nu-\frac{A-1}{2A}}{1-\nu-\frac{A-1}{2A}}.\label{eq:A12}
\end{equation}
In the case of elastically isotropic crystal, one has the Lamé coefficients
$\lambda=c_{12}$ and $\mu=c_{44}=(c_{11}-c_{12})/2$, the Poisson
ratio being $\nu=\lambda/2(\lambda+\mu)$. Then, $s_{c}=0$ and, calculating
the coefficient $\alpha$ by Eq.~(\ref{eq:A9}), we get $\alpha=\nu/(1-\nu)$.

The elastic constants of diamond give $\nu=0.103$ and $(A-1)/2A=0.087$.
Both quantities are small, but not exceptionally small. However, the
coefficient $\alpha$ is given by the difference between the Poisson
and the anisotropy parameters and occurs numerically exceptionally
small. For a comparison, the elastic constants of silicon give $\nu=0.278$
and $(A-1)/2A=0.180$, so that the Poisson and the anisotropy effects
only partially compensate each other.

\section{Components of the scattering vector}

\label{sec:AppendixB}

The aim of this Appendix is to derive explicit expressions for the
components of the deviation $\mathbf{q}$ of the scattering vector
from the reciprocal lattice vector $\mathbf{Q}$, taking into account
both an angular deviation of the incident beam $\delta\theta$ from
Bragg orientation and a wave vector deviation $\delta k$ from the
reference wave vector $k_{0}$. We introduce the wave vectors $\mathbf{K}_{0}^{\mathrm{in}}$
and $\mathbf{K}_{0}^{\mathrm{out}}$, satisfying the Bragg law for
the reference wave length, $\mathbf{K}_{0}^{\mathrm{out}}-\mathbf{K}_{0}^{\mathrm{in}}=\mathbf{Q}$
and $\left|\mathbf{K}_{0}^{\mathrm{out}}\right|=\left|\mathbf{K}_{0}^{\mathrm{in}}\right|=k_{0}$.
The wave vector of the incident wave ${\mathbf{K}}^{\mathrm{in}}=k(\cos\theta^{\mathrm{in}},\sin\theta^{\mathrm{in}})$
differs from the reference wave vector ${\mathbf{K}}_{0}^{\mathrm{in}}=k_{0}(\cos\theta_{B},\sin\theta_{B})$
due to both an incidence angle deviation $\theta^{\mathrm{in}}=\theta_{B}+\delta\theta$
and a deviation of the wave vector length $k=k_{0}+\delta k$. Hence,
the difference $\mathbf{q}^{\mathrm{in}}=\mathbf{K}^{\mathrm{in}}-\mathbf{K}_{0}^{\mathrm{in}}$
is equal to 
\begin{equation}
{\mathbf{q}}^{\mathrm{in}}=\left(\delta k\cos\theta_{B}-k_{0}\delta\theta\sin\theta_{B},\,\delta k\sin\theta_{B}+k_{0}\delta\theta\cos\theta_{B}\right).\label{eq:K7}
\end{equation}
Similarly, for the wave vector of the scattered wave in symmetric
Bragg case ${\mathbf{K}}^{\mathrm{out}}=k(\cos\theta^{\mathrm{out}},-\sin\theta^{\mathrm{out}})$
with its angular deviation from the reference beam direction $\theta^{\mathrm{out}}=\theta_{B}+\delta\theta'$
and the same wave vector as the incident beam $k=k_{0}+\delta k$,
the difference $\mathbf{q}^{\mathrm{out}}=\mathbf{K}^{\mathrm{out}}-\mathbf{K}_{0}^{\mathrm{out}}$
is 
\begin{equation}
{\mathbf{q}}^{\mathrm{out}}=\left(\delta k\cos\theta_{B}-k_{0}\delta\theta'\sin\theta_{B},\,-\delta k\sin\theta_{B}-k_{0}\delta\theta'\cos\theta_{B}\right).\label{eq:K8}
\end{equation}
The components of the wave vector $\mathbf{q}=\mathbf{K}^{\mathrm{out}}-\mathbf{K}^{\mathrm{in}}-\mathbf{Q}=\mathbf{q}^{\mathrm{out}}-\mathbf{q}^{\mathrm{in}}$
are 
\begin{eqnarray}
q_{x} & = & k_{0}(\delta\theta-\delta\theta')\sin\theta_{B},\nonumber \\
q_{z} & = & -2\delta k\sin\theta_{B}-k_{0}(\delta\theta+\delta\theta')\cos\theta_{B}.\label{eq:K9}
\end{eqnarray}

It is convenient to represent the scattered intensity in an energy
spectrum, considering the scattering angle $2\theta_{B}+\delta\theta+\delta\theta'$
as twice the Bragg angle for the respective wavelength, i.e. as $\delta k'/k_{0}=-(\delta\theta+\delta\theta')/(2\tan\theta_{B})$.
Then, the components of the wave vector $\mathbf{q}$ are 
\begin{eqnarray}
q_{x} & = & 2(k_{0}\delta\theta-\delta k'\tan\theta_{B})\sin\theta_{B},\nonumber \\
q_{z} & = & 2(\delta k'-\delta k)\sin\theta_{B}.\label{eq:K12}
\end{eqnarray}

\section{Kinematical scattering amplitude for a finite width of the incident
beam}

\label{sec:AppendixC}

Consider a Gaussian spacial distribution of the amplitude of the wave
incident on the bent crystal, 
\begin{equation}
A_{0}(x,z)=\exp\left(-4\xi^{2}/w^{2}\right),\label{eq:1}
\end{equation}
where $\xi=x\sin\theta_{B}-z\cos\theta_{B}$ is the distance in the
direction normal to the incidence beam. This term has to be included
in the integrand of Eq.~(\ref{eq:K3}), so that the amplitude of
the diffracted wave for a spatially limited incidence beam can be
written as

\begin{eqnarray}
A(q_{x},q_{z}) & = & \intop_{-\infty}^{\infty}dx\intop_{-D/2}^{D/2}dz\,e^{-i(q_{x}x+q_{z}z)}\label{eq:K4}\\
 &  & \times e^{-4(x\sin\theta_{B}-z\cos\theta_{B})^{2}/w^{2}}e^{iQ(x^{2}+\alpha z^{2})/2R}.\nonumber 
\end{eqnarray}
This integral can be expressed through the Faddeeva function of complex
argument $W(z)=\exp(-z^{2})\mathrm{\,erfc}(-iz)$, where $\mathrm{erfc}(z)$
is the complimentary error function. Free codes to evaluate $W(z)$
are available \cite{poppe90,weideman94}. Calculation of the integral
gives 
\begin{eqnarray}
A(q_{x},q_{z}) & = & f\exp\left(iq_{x}^{2}R^{2}w^{2}/2l^{2}\right)\nonumber \\
 & \times & \left[\exp\left(-i\frac{pD}{2l^{2}}\right)W\left(-\frac{2p-ig^{2}D}{2\sqrt{2}gl}\right)\right.\label{eq:K5}\\
 &  & -\left.\exp\left(i\frac{pD}{2l^{2}}\right)W\left(-\frac{2p+ig^{2}D}{2\sqrt{2}gl}\right)\right],\nonumber 
\end{eqnarray}
where two complex parameters which have dimension of length 
\begin{eqnarray}
l & = & \sqrt{-QRw^{2}-8iR^{2}\sin^{2}\theta_{B}},\nonumber \\
p & = & q_{z}l^{2}-4iq_{x}R^{2}\sin2\theta_{B},\label{eq:K6}
\end{eqnarray}
and two complex dimensionless parameters 
\begin{eqnarray}
g & = & \sqrt{-8QR\left(\cos^{2}\theta_{B}+\alpha\sin^{2}\theta_{B}\right)+i\alpha(Qw)^{2}},\nonumber \\
f & = & \frac{(1-i)\pi Rw}{\sqrt{2}g}\label{eq:K10}\\
 &  & \times\exp\left[\frac{QRD^{2}}{l^{2}}-i\frac{QD^{2}[Qw^{2}+(1-\alpha)l^{2}R]}{8l^{2}}\right]\nonumber 
\end{eqnarray}
are introduced.

\referencelist[surface]
\end{document}